\documentclass[aps,prl,showpacs,twocolumn,floats]{revtex4}
\usepackage{graphicx}
\usepackage{dcolumn}
\usepackage{bm}

\begin{document}

\title{Rotational States of Magnetic Molecules}
\date{\today}
\author{E. M. Chudnovsky and D. A. Garanin}
\affiliation{Physics Department, Lehman College, City University
of New York \\ 250 Bedford Park Boulevard West, Bronx, New York
10468-1589, USA}
\date{\today}

\begin{abstract}
We study a magnetic molecule that exhibits spin tunneling and is
free to rotate about its anisotropy axis. Exact low-energy
eigenstates of the molecule that are superpositions of spin and
rotational states are obtained. We show that parameter $\alpha =
2(\hbar S)^2/(I\Delta)$ determines the ground state of the
molecule. Here $\hbar S$ is the spin, $I$ is the moment of
inertia, and $\Delta$ is the tunnel splitting. The magnetic moment
of the molecule is zero at $\alpha < \alpha_c =
[1-1/(2S)^{2}]^{-1}$ and non-zero at $\alpha
> \alpha_c$. At $\alpha \rightarrow \infty$ the spin of the molecule
localizes in one of the directions along the anisotropy axis.
\end{abstract}
 \maketitle

Crystals of high-spin magnetic molecules came to the attention of
physicists after Sessoli et al. \cite{Sessoli} discovered that
they behave as regular arrays of identical superparamagnetic
particles \cite{Lectures}. The remarkable property of magnetic
molecules is that their spin can tunnel between up and down
directions \cite{MQT-book}. This leads to a characteristic
step-wise magnetization curve discovered by Friedman et al. in
Mn$_{12}$-Acetate \cite{Friedman} and later observed in hundreds
of other molecular magnets. More recently, experiments were
performed with magnetic molecules deposited on surfaces
\cite{surfaces} and with single molecules bridged between metallic
electrodes \cite{electrodes}. The interest to such studies has
been driven in part by the prospect of using magnetic molecules as
qubits \cite{qubits}. At first glance, partial or total decoupling
of the molecule from the environment appears desirable to achieve
low decoherence. It was noticed \cite{EC-PRL94,XM-EC}, however,
that such a decoupling may prohibit spin tunneling altogether due
to conservation of the total angular momentum, ${\bf J} = {\bf S}
+ {\bf L}$, with ${\bf S}$ being the spin of the molecule and
${\bf L}$ being the orbital angular momentum associated with the
mechanical rotation. This situation can be relevant to recent
experiments with single magnetic molecules if the molecules
maintain some degree of freedom with respect to rotations.

Consider first a high-spin molecule in a crystal. The general form
of the spin Hamiltonian of the molecule is
\begin{equation}\label{ham}
\hat{H}_S = \hat{H}_{\parallel} + \hat{H}_{\perp}\,,
\end{equation}
where $\hat{H}_{\parallel}$ commutes with $S_z$ and
$\hat{H}_{\perp}$ is a perturbation that does not commute with
$S_z$. The existence of the magnetic anisotropy axis means that
the $| \pm S\rangle$ eigenstates of $S_z$ are degenerate ground
states of $\hat{H}_{\parallel}$. Operator $\hat{H}_{\perp}$
slightly perturbs the $| \pm S\rangle$ states, adding to them
small contributions of other $|m_S\rangle$ states. We shall call
these degenerate normalized perturbed states $|\psi_{\pm
S}\rangle$. Physically they describe the magnetic moment of the
molecule looking in one of the two directions along the anisotropy
axis. Full perturbation theory with account of the degeneracy of
$\hat{H}_{S}$ provides quantum tunneling between the $|\psi_{\pm
S}\rangle$ states. The ground state and the first excited state
become
\begin{equation}\label{pm}
\Psi_{\pm} = \frac{1}{\sqrt{2}}\left(|\psi_{S}\rangle \pm
|\psi_{-S}\rangle\right)\,.
\end{equation}
They satisfy
\begin{equation}\label{Epm}
\hat{H}_S\Psi_{\pm} = E_{\pm}\Psi_{\pm}
\end{equation}
with $E_- - E_+ \equiv \Delta$ being the tunnel splitting. Since
the crystal field Hamiltonian $\hat{H}_S$ does not possess the
full invariance with respect to rotations, $\Psi_{\pm}$ should not
be the eigenstates of ${\bf J}$. However, a closed system
consisting of the spin and the crystal does possess such
invariance. It has been demonstrated \cite{EC-universal,CGS} that
conservation of the total angular momentum (spin + crystal)
dictates entanglement of spin states with elastic twists. This
effect contributes to spin decoherence but does not significantly
affect the ground state energy.

The situation changes for a free magnetic molecule. A high-spin
molecule usually consists of hundreds of atoms, making its
mechanical properties similar to the mechanical properties of a
tiny solid body. Free magnetic clusters in beams have been studied
in the past \cite{beams}. They exhibit a number of interesting
phenomena some of which have been attributed to the interaction
between spin and mechanical degrees of freedom. General analytical
solution for the rotational quantum levels of a rigid body does
not exist. Spin degree of freedom further complicates the problem.
However, as we demonstrate below, the exact eigenstates and exact
energy levels can be obtained analytically for the low-energy
states of a magnetic molecule that is free to rotate about its
anisotropy axis. This could be the case when a free molecule is in
a magnetic field or the molecule is bridged between two leads. The
eigenstates of such a molecule must be the eigenstates of $J_z =
S_z + L_z$. It is then clear that, unless mechanical rotations are
involved, conservation of $J_z$ prohibits quantum tunneling of
${\bf S}$. For, e.g., $J_z =0$ the transitions can only occur
between the states $|\psi_{S}\rangle \otimes |m_L = -S\rangle$ and
$|\psi_{-S}\rangle \otimes |m_L = S\rangle$. These are the states
in which the angular momentum due to spin is compensated by the
angular momentum due to mechanical rotation. For a superposition
of these states to be the ground state of the system, the kinetic
energy, $(\hbar S)^2/(2I)$, associated with the rotation cannot
significantly exceed the energy gain, $\Delta/2$, due to spin
tunneling. Otherwise the ground state will be $|\psi_{\pm
S}\rangle \otimes |m_L = 0\rangle$. For a solid particle, the
moment of inertia $I$ grows as the fifth power of the size of the
particle. Consequently, rotational effects should be less
important in large particles. For magnetic molecules, however, the
rotational energy $(\hbar S)^2/(2I)$ in many cases will be large
enough to cause localization in one of the $|\psi_{\pm S}\rangle$
spin states. Exact analytical solution of this problem is given
below.

Since the low-energy spin states of the molecule are
superpositions of $|\psi_{\pm S}\rangle$, it is convenient to
describe such a two-state system by a pseudospin 1/2. Components
of the corresponding Pauli operator ${\bm \sigma}$ are
\begin{eqnarray}\label{Pauli}
\sigma_x & = & |\psi_{-S}\rangle\langle \psi_{S}| +
|\psi_{S}\rangle\langle \psi_{-S}|
\\ \nonumber \sigma_y & = &
i|\psi_{-S}\rangle\langle \psi_{S}| - i|\psi_{S}\rangle\langle \psi_{-S}| \\
\nonumber \sigma_z & = & |\psi_{S}\rangle\langle \psi_{S}| -
|-\psi_{S}\rangle\langle \psi_{-S}|\,.
\end{eqnarray}
The projection of $\hat{H}_S$ onto $|\psi_{\pm S}\rangle$ states
is
\begin{equation}\label{projection}
\hat{H}_{\sigma} = \sum_{m,n = \psi_{\pm S}}\langle
m|\hat{H}_S|n\rangle|m\rangle\langle n|\,.
\end{equation}
Expressing $|\psi_{\pm S}\rangle$ via $\Psi_{\pm}$ according to
Eq.\ (\ref{pm}), it is easy to see from Eq.\ (\ref{Epm}) that
\begin{equation}\label{ME}
\langle \psi_{\pm S}|\hat{H}_S|\psi_{\pm S}\rangle =  0, \quad
\langle \psi_{-S}| \hat{H}_S|\psi_{S}\rangle = - {\Delta}/{2}\,.
\end{equation}
With the help of these relations one obtains from Eq.\
(\ref{projection})
\begin{equation}\label{projection1}
\hat{H}_{\sigma} = -\frac{\Delta}{2}\sigma_x\,.
\end{equation}

So far we have not considered mechanical rotations of the
molecule. Rotation by angle $\phi$ about the anisotropy axis $Z$,
transforms the spin Hamiltonian into
\begin{equation}\label{spinrotation}
\hat{H}_S' = e^{-iS_z\phi}\hat{H}_S e^{iS_z\phi}\,.
\end{equation}
Noticing that
\begin{equation}
S_z|\psi_{\pm S}\rangle \cong S_z|\pm S\rangle = \pm S |\psi_{\pm
S}\rangle\,,
\end{equation}
it is easy to project Hamiltonian (\ref{spinrotation}) onto
$\psi_{\pm S}$. Simple calculation yields the following
generalization of Eq.\ (\ref{projection1}):
\begin{eqnarray}\label{projection'}
\hat{H}_{\sigma}' & = & \sum_{m,n = \psi_{\pm S}}\langle
m|\hat{H}_S'|n\rangle|m\rangle\langle n| \\ \nonumber & = &
-\frac{\Delta}{2}\left[\cos(2S\phi)\sigma_x +
\sin(2S\phi)\sigma_y\right]\,.
\end{eqnarray}
The full Hamiltonian of the magnetic molecule rotating about its
anisotropy axis is
\begin{equation}\label{H-phi}
\hat{H} = \frac{(\hbar L_z)^2}{2I} -\frac{\Delta}{2}\left[
\sigma_x \cos(2S\phi)+ \sigma_y \sin(2S\phi)\right]\,,
\end{equation}
with $L_z = -i(d/{d \phi})$.

We are now in a position to find the eigenstates of the rotating
molecule. By construction, the Hamiltonian (\ref{H-phi}) is
invariant with respect to rotations about the $Z$-axis.
Consequently, its eigenstates must be the eigenstates of $J_z =
L_z + S_z$:
\begin{equation}\label{Psi-J}
\Psi_J =
\frac{1}{\sqrt{2}}\left(C_S|\psi_{S}\rangle\otimes|J-S\rangle_l +
C_{-S}|\psi_{-S}\rangle\otimes|J +S\rangle_l\right) \, .
\end{equation}
Here $J \equiv m_J$ while index $l$ denotes states in the
mechanical space, with $|m \rangle_l \equiv |m_L\rangle =
\exp(im_L\phi)$. Solution of $\hat{H}\Psi_J = E_J\Psi_J$ gives the
following expression for the energy levels:
\begin{equation}\label{levels}
E_{J\pm} = \frac{\Delta}{2}\left[\left(1 +
\frac{J^2}{S^2}\right)\frac{\alpha}{2} \pm \sqrt{1 +
\frac{J^2}{S^2}\, \alpha^2}\right]\,,
\end{equation}
where
\begin{equation}
\alpha \equiv \frac{2(\hbar S)^2}{I\Delta}\,.
\end{equation}
For $J \neq 0$ each state is degenerate with respect to the sign
of $J$. For $J = 0, 1, 2, ...$ coefficients $C_{\pm}$ are given by
\begin{eqnarray}\label{C}
C_{S} & = & \sqrt{1 + \frac{\alpha J}{\sqrt{S^2 + (\alpha J)^2}}}
\nonumber \\
C_{-S} & = & \mp\sqrt{1 - \frac{\alpha J}{\sqrt{S^2 + (\alpha
J)^2}}}\,,
\end{eqnarray}
where $\mp$ correlates with $\pm$ in Eq.\ (\ref{levels}).

\begin{figure}[ht]
\begin{center}
\includegraphics[width=60mm,angle=-90]{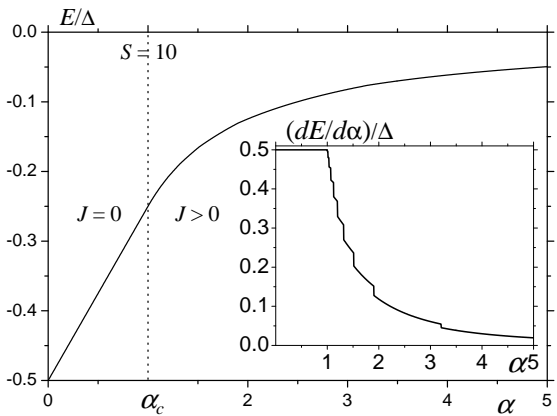}
\caption{Zero-field ground state energy as a function of $\alpha$.
Inset shows the derivative of the ground state energy on
$\alpha$.} \label{GS-energy}
\end{center}
\end{figure}

At $J \approx m_L \gg S$, Eq.\ (\ref{levels}) gives the energy of
the mechanical rotation, $(\hbar m_L)^2/(2I)$. At small $\alpha$
the ground state and the first excited state correspond to $J =
0$,
\begin{equation}\label{small-alpha}
E_{0\pm} = \frac{\hbar^2S^2}{2I} \pm \frac{\Delta}{2}\,.
\end{equation}
Here the first term is the energy of the rotation with $m_L = \pm
S$. For a molecule rigidly coupled to an infinite mass one has $I
\rightarrow \infty$ and the energy of the rotation goes to zero.
In this case one recovers from Eq.\ (\ref{small-alpha}) the
energies, $\pm \Delta/2$, of the tunnel-split spin states in a
macroscopic crystal. As $\alpha$ increases, the ground state
switches to higher $J$. The value of $\alpha$ at which the ground
state changes from $E_{J-1}$ to $E_{J}$ satisfies
\begin{equation}\label{GS-J}
E_{J-1,-}(\alpha_{J}) = E_{J,-}(\alpha_{J})\,.
\end{equation}
Solution of this equation for $J = 1, 2, ..., S$ gives
\begin{equation}\label{critical-alpha}
\alpha_{J} = \left[1 - \frac{1}{(2S)^2}\right]^{-1/2}\left[1 -
\frac{(2J-1)^2}{(2S)^2}\right]^{-1/2}\,.
\end{equation}
For $\alpha$ smaller then
\begin{equation}
\alpha_1 = \left[1 - 1/(2S)^2\right]^{-1}
\end{equation}
the ground state corresponds to $J = 0$, $C_{\pm S} = 1$. At
$\alpha = \alpha_1$ the transition to the $J=1$ ground state takes
place. At $\alpha = \alpha_2$ the ground state changes from $J =
1$ to $J=2$,  and so on. At $\alpha$ greater than
\begin{equation}
\alpha_{S} = \left[1 - \frac{1}{(2S)^2}\right]^{-1/2}\left[1 -
\left(1 -\frac{1}{2S}\right)^2\right]^{-1/2}
\end{equation}
the ground state always corresponds to $J = S$. For, e.g., $S =
10$ one obtains $\alpha_1 = 1.0025$ and $\alpha_{10} = 3.2066$.
The dependence of the ground state energy on $\alpha$ for $S=10$
is shown in Fig.\ \ref{GS-energy}. While this dependence is
smooth, the derivative of the ground state energy on $\alpha$
shows steps at the critical values of $\alpha$ given by Eq.\
(\ref{critical-alpha}). In the limit of $\alpha \rightarrow 0$ the
ground state energy is $-\Delta/2$. This is the gain in energy due
to spin tunneling between $|\psi_{\pm S}\rangle$ states in an
infinitely heavy particle. In the limit of $\alpha \gg 1$ (light
particle) $J = S$ and according to Eq.\ (\ref{levels}) the ground
state energy approaches zero as $-\Delta/(4\alpha)$. This
corresponds to the gradual localization of the spin in one of the
$|\psi_{\pm S}\rangle$ states.

\begin{figure}[ht]
\begin{center}
\includegraphics[width=60mm,angle=-90]{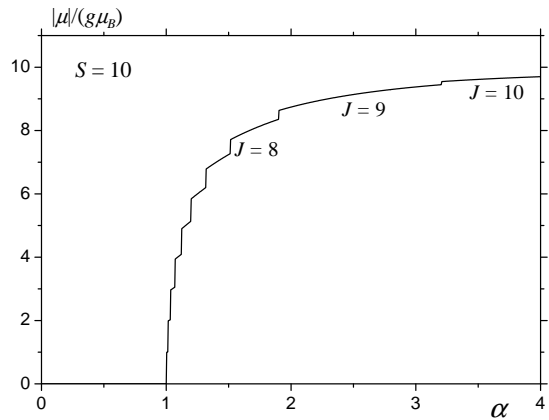}
\caption{Dependence of the ground-state magnetic moment on
parameter $\alpha$.} \label{moment}
\end{center}
\end{figure}

To compute the magnetic moment of the molecule we notice that
$L_z$ in our formalism describes the mechanical rotation of the
molecule as a whole, not the orbital states of the electrons.
Consequently, the magnetic moment of a free magnetic molecule
should be entirely due to its spin:
\begin{equation}\label{mag-moment}
\mu  =  -g\mu_B\langle \Psi_J|S_z|\Psi_J\rangle  =  -g\mu_B
S\frac{\alpha J}{\sqrt{S^2 + (\alpha J)^2}}\,.
\end{equation}
Here $g$ is the spin gyromagnetic factor. The minus sign reflects
negative gyromagnetic ratio, $\gamma = - g\mu_B/\hbar$, for the
electron spin. If the ground state corresponds to $ J = 0$,
spin-up and spin-down states contribute equally to the wave
function and the magnetic moment is zero. When $J$ in the
ground-state is non-zero, spin-up and spin-down states contribute
with different weights and the molecule has a non-zero magnetic
moment. Which $J$ corresponds to the ground state depends on the
parameter $\alpha$. The dependence of the ground state magnetic
moment on $\alpha$ is shown in Fig.\ \ref{moment}.

The above results can be easily generalized to take into account
the effect of the external magnetic field $B$ applied along the
$Z$-axis. Such a field adds a Zeeman term, $g\mu_B S_z B$, to the
Hamiltonian (\ref{ham}). This term is invariant with respect to
the rotation by the angle $\phi$. Its projection onto $\psi_{\pm
S}$ simply adds $g\mu_B S B \sigma_z$ to Eq.\ (\ref{H-phi}). The
full projected Hamiltonian becomes
\begin{equation}\label{ham-B}
\hat{H} = -\frac{\hbar^2}{2I}\frac{d^2}{d\phi^2}
-\frac{\Delta}{2}\left[ \sigma_x \cos(2S\phi)+ \sigma_y
\sin(2S\phi)\right] - \frac{W}{2}\sigma_z\,,
\end{equation}
where
\begin{equation}\label{W}
W \equiv -2g \mu_B S B\,.
\end{equation}
Since this Hamiltonian is invariant with respect to rotations
about the $Z$-axis, its eigenfunctions are still given by Eq.\
(\ref{Psi-J}) with the coefficients $C_{\pm S}$ now depending on
$B$. Solving $\hat{H}\Psi_J = E_J\Psi_J$ one obtains
\begin{equation}\label{levels-B}
E_{J\pm} = \frac{\Delta}{2}\left[\left(1 +
\frac{J^2}{S^2}\right)\frac{\alpha}{2} \pm \sqrt{1 +
\left(\frac{W}{\Delta} + \frac{J}{S}\, \alpha\right)^2}\right]
\end{equation}
for the energy levels. Here $W$ can be positive or negative
depending on the orientation of the field. Positive $W$
corresponds to the magnetic field in the direction of the magnetic
moment, which provides the lower energy. At $B \neq 0$
coefficients $C_{\pm S}$ can be presented in the form
\begin{equation}\label{C-B}
C_{S}  =  \sqrt{1 + \frac{\bar{W}}{\sqrt{\Delta^2 +
\bar{W}^2}}}\,, \quad C_{-S} =  \mp\sqrt{1 -
\frac{\bar{W}}{\sqrt{\Delta^2 + \bar{W}^2}}}\,,
\end{equation}
with
\begin{equation}\label{W-bar}
\bar{W} \equiv  W + \alpha J \Delta/S \,.
\end{equation}
Notice that Eq.\ (\ref{C-B}) coincides with the form of $C_{\pm
S}$ for a frozen magnetic molecule in the magnetic field $\bar{B}
= B - \hbar J/(\gamma I)$. The magnetic moment of the molecule
that is free to rotate is given by
\begin{equation}\label{mag-moment-B}
\mu  =  -g\mu_B S\frac{\bar{W}}{\sqrt{\Delta^2 + \bar{W}^2}}\,.
\end{equation}

\begin{figure}[ht]
\begin{center}
\includegraphics[width=60mm,angle=-90]{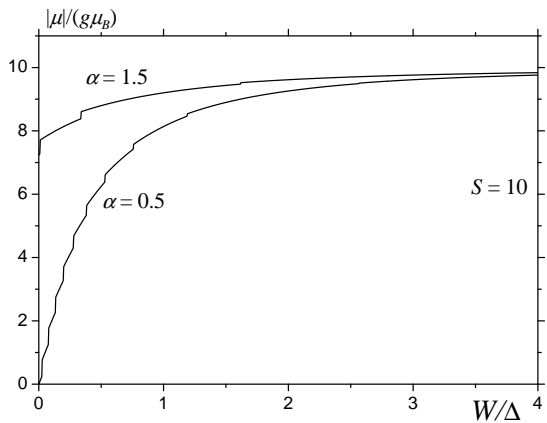}
\caption{Field dependence of the magnetic moment. Note the jumps
at $W = W_J$.} \label{moment-B}
\end{center}
\end{figure}

In the absence of the magnetic field, quantum number $J$
corresponding to the ground state is determined by $\alpha$. For a
given magnetic molecule this parameter is fixed. On the contrary,
in the presence of the field, $J$ can be manipulated by changing
$B$. Solving Eq.\ (\ref{GS-J}) with $E_{J\pm}$ of Eq\
(\ref{levels-B}), one obtains the following expression for $W =
W_J $ at which the ground state switches from $J-1$ to $J$:
\begin{equation}\label{W-J}
\frac{W_J}{\Delta} = \frac{2J - 1}{2S}\left\{\sqrt{\left[1 -
\left(\frac{2J - 1}{2S}\right)^2\right]^{-1} +
\left(\frac{\alpha}{2S}\right)^2} - \alpha\right\}.
\end{equation}
Here $J = 1, 2, ..., S$. The dependence of the ground state
magnetic moment on $W$ is shown in Fig.\ \ref{moment-B}.
The jumps at critical values of the field must show as sharp
maxima in the differential susceptibility.

In conclusion, we have obtained exact low-energy quantum states of
a magnetic particle (molecule) that exhibits spin tunneling and is
free to rotate about its anisotropy axis. The ground state depends
on the parameter $\alpha = 2(\hbar S)^2/(I\Delta)$. Various limits
studied above are physically accessible in magnetic molecules and
atomic clusters. At $\alpha \rightarrow \infty$ the spin localizes
in one of the two directions along the magnetic anisotropy axis.
Magnetic molecule of a nanometer size has the moment of inertia in
the ballpark of $10^{-35}$ g cm$^2$. For $S = 10$ this provides
$\alpha \sim 1$ at $\Delta/\hbar \sim 10^{10}$s$^{-1}$. The tunnel
splitting of, e.g., Mn$_{12}$ and Fe$_8$ molecules is much
smaller. Thus the spin tunneling in these molecules must be
strongly suppressed if they are free to rotate. This effect may be
important in designing qubits based upon magnetic molecules.

Authors acknowledge support from the National Science Foundation
through Grant No. DMR-0703639.

\end{document}